# The use of the open innovation paradigm in the public sector: a systematic review of published studies

Joel Alves de Lima Júnior
jalj@cin.ufpe.br

Center of Informatics, Federal University of Pernambuco, Brazil

Kiev da Gama Santos
kiev@cin.ufpe.br

Center of Informatics, Federal University of Pernambuco, Brazil

Jorge da Silva Correia Neto
jorgecorreianeto@gmail.com

Academic University of Distance Education and Technology, Federal Rural University of Pernambuco, Brazil

**Abstract**

The use of the open innovation paradigm has been, over the past years, getting special attention in the public sector. Motivated by an urban environment that is increasingly more complex and challenging, several government agencies have been allocating financial resources and efforts to promote open and participative government initiatives. As a way to try and understand this scenario, a systematic review of the literature was conducted, to provide a comprehensive analysis of the scientific papers that were published, seeking to capture, classify, evaluate and synthesize how the use of this paradigm has been put into practice in the public sector. In total, 4,741 preliminary studies were analyzed. From this number, only 37 articles were classified as potentially relevant and moved forward, going through the process of data extraction and analysis. From the data obtained, it was possible to verify that the use of this paradigm started to be reported with a higher frequency in the literature since 2013 and, among the main findings, we highlight the reports of experiences, approach propositions, of understanding how the phenomenon occurs and theoretical reflections. It was also possible to verify that the use of open innovation through social media was one of the pioneer techniques of engagement between the public sector and citizens. In conclusion, the reports confirm that the main challenges of this paradigm applied to the public sector are associated with their respective bureaucratic aspects, therefore lacking a bigger reflection on the procedures and methods to be used in the public sphere.

## 1 Introduction

Open Innovation (OI) is a term that was originally conceived in the field of business strategy and innovation (Seltzer & Mahmoudi, 2013), being introduced by Chesbrough (Chesbrough, Press, & Brown, 2003) as a paradigm shift that converged from the traditional model of closed innovation, about the internal rigid control of ideas and knowledge resources inside an organization, to a new open model that seeks the effort of external agents as a way to incorporate ideas originated outside of the organization towards innovation processes that take place inside the company environment.

With a high increase of successful cases in the private sector, public organizations also started to adopt innovation processes aiming to expand the participation of citizens in the public sector and promote improvement in the quality of public services provided. Gascó (2017) underlines that to enhance OI in the public sector, Information and Communication Technologies (ICT) contribute a lot in this process.

Although extensive research on adopting OI in the private sector has already been carried out, fundamental differences in its implementation present a series of challenges that demand a deeper investigation. Regarding the innovation field in the public sector (Androutsopoulou, Karacapilidis, Loukis, & Charalabidis, 2017), contrary to the private sector which has the goal to increase profit, the public sector aims to increase benefits for those who are interested and to improve the quality of the public services provided (Yang & Kankanhalli, 2013). Thus, the importance of developing studies that clarify how public institutions innovate by adopting the open innovation paradigm comes to light. This helps to increase the understanding of this phenomenon applied to the public sector that, as pointed out by the literature, is still depending on results that demonstrate its applicability in that sector.

This study presents a systematic review that aims to analyze models, methods and approaches of how the public sector has been assimilating the open innovation paradigm to promote innovation, investigating the facilitating aspects and identifying current gaps. With this in mind, we designed the following research question:

- How does the open innovation process take place in the public sector?

From this research question, other specific questions were generated, as enumerated below:

RQ1 What open innovation models for governments are there in the literature?
RQ2 What are the approaches used for open innovation in the public sector?
RQ3 Which stakeholders are involved in these initiatives in the public sector?
RQ4 What aspects influence open innovation in the public sector?
RQ5 What is the role of ICTs in open innovation processes in the public sector?

To answer these questions, a research was conducted on the primary studies published from 2009 to 2020. We chose 2009 as the starting date due to the publication of the US Government memo regarding data transparency and its availability in an open format (The White House, 2009). The search process resulted in a total of 4,741 articles, from which 37 were relevant and selected for the process of analysis and extraction of data. Therefore, the goal of this article is to discuss the innovation approaches that are being proposed and applied in the public sector throughout these years, investigating the facilitating processes of these innovation practices and identifying current gaps. The contribution of this study is to reflect on how this paradigm has been incorporated in different public organizations throughout the years.

This article is organized as follows. In section 2 we established the concepts that set the ground to the analysis that was carried out. In section 3 we presented the main elements of the literature method of review that was used. In section 4 the results of the review are presented, answering the five research questions presented above. In section 5, we conducted a discussion regarding the obtained results, future works and final considerations.

## 2 Background

### 2.1 Open Innovation

Open Innovation is a concept that was initially conceived in the business strategy and innovation field. It is referred to the rational effort from companies to incorporate ideas that originated outside the companies, into innovation processes, towards the institution, or to promote internal ideas for a commercial application (Seltzer & Mahmoudi, 2013). Differently from the traditional closed innovation model, where innovation was developed and disseminated by organizations without the cooperation from third parties (Chesbrough, Vanhaverbeke, & West, 2006), open innovation seeks the engagement of users with the purpose of thinking of innovation as a way to increase awareness inside companies or organizations.

The open innovation model was originally presented by Henry Chesbrough (Chesbrough et al., 2003) and, based on the author, the concept of openness is related to the idea that innovation can't be achieved in an isolated manner since there is a dependency on several partners in order to collect ideas and resources. Among the several meanings, open innovation may be understood as a process where knowledge flows through organizational frontiers (West, Salter, Vanhaverbeke, & Chesbrough, 2014). These flows can be sorted by two classifications: 1) inbound - imported from outside into the organization, thus accelerating the internal development through sources of knowledge that are acquired externally, and; 2) outbound - from inside towards outside the organization, providing ideas and technologies through intellectual property, licenses or patents, developed internally, to share with external agents (Chesbrough & Crowther, 2006).

Open innovation consists of an emergent model of innovation that incorporates knowledge derived from external and internal sources, taking into consideration that companies can and should use external and internal paths to the market while seeking to improve their technology (Chesbrough, Vanhaverbeke, & West, 2014). It is an approach associated with opening an organization process, aiming to exchange experiences, ideas and knowledge with partners, clients and/or suppliers in their innovation and competitive strategy processes (Enkel, Gassmann, & Chesbrough, 2009). One of the aspects that motivate this openness is that companies that only focus internally are at risk of losing opportunities since most opportunities are outside the scope of their current internal activities and they need to connect with external technologies to unlock their potential (Chesbrough et al., 2003).

In the literature, open innovation consists of different aspects concerning the role of the external user included in the innovation process, which has several different names, such as user-guided innovation (von Hippel, 2005), user engagement (Magnusson, 2003), or co-creation (Prahalad & Ramaswamy, 2004). Despite different ways to understand it, there is a common denominator: final users are engaged in innovation or development processes as active participants instead of passive ones (Dan Breznitz & Rouvinen, 2009).

## 2.2 Open Innovation in Public Sector

Open innovation boosts internal innovation teams with knowledge spread around the world. As the cornerstone for the future development of any nation, the open innovation paradigm was expanded in all levels of government (Georghiou, Edler, Uyarra, & Yeow, 2014), as many governments around the globe started to design strategies, develop innovation structures and link outcomes to their National Agenda. Consequently, each government entity uses its own basic process to enable innovation in its work environment.

The use of open innovation in the public sector is frequently associated with the possibility of expanding citizen participation in the public sector, which leads to the improvement or creation of new public services (Barton Cunningham & Kempling, 2009). The US Federal Government made in 2009 an important commitment to the Open Government Initiative, allowing citizens to access government data and to contribute with ideas and knowledge to formulate governmental policies and services of innovation (The White House, 2009). An important advance in international terms occurred in 2011, with the creation of the Open Government Partnership (OGP). This organization was created aiming to globally disseminate and encourage good governmental practices, access to information, the combat against corruption, social participation and the use of new technologies to innovate government and strengthen governance, stimulating countries to adopt practices that promote a more open, effective and accountable public management (Open Government Partnership, 2022).

Because of that, several countries in the world started to adopt practices to make data available in an open format and to expand mechanisms that enable innovation along-side citizens. In this regard, it is possible to mention the Open Data Portal of the United Kingdom (https://www.data.gov.uk/), which currently hosts 27,742 open databases clustered in 14 specific categories. The US Government Data.Gov (https://data.gov) currently hosts a total of 20,977 public data from a total of 153 agencies involved. The Singapore Government Portal (https://data.gov.sg), created in 2011, currently has a number of 1,877 open databases. The De Publieke Zaak (http://publiekezaak.nl) from the Netherlands enables government agencies to innovate using citizens' insights. This has also been explored by developing countries, and it is possible to mention the Brazilian Portal of Open Data (https://dados.gov.br), which hosts a total of 12,969 combined data, relying on the engagement of 209 government organizations. The open innovation initiatives started being adopted in several countries, as an initial effort to promote open practices. The availability of open data is a prior requirement for successful open innovation activities (Thoreson & Miller, 2013).

In the past decade, the open innovation paradigm has become well known among researchers and professionals in regard to its use in the private sector (Biscotti, Mafrolla,

Giudice, & D'Amico, 2018; Bogers et al., 2016). Motivated by these trends and by the rise of successful cases of this paradigm in the private sector (Gascó, 2017), a growing number of public institutions have been adopting this strategy (Bommert, 2010; Georghiou et al., 2014) with the objective to engage citizens in the public aspects (Barton Cunningham & Kempling, 2009; Ferraris, Belyaeva, & Bresciani, 2020), boost collective intelligence, design products and solve issues (Brabham, 2013).

Particularly, the closed innovation model doesn't address the policy challenges that the public sector organizations have to deal with, thus justifying the need for the public sector to adopt the open innovation model (Bommert, 2010). With the increase of citizen dissatisfaction towards the State and because they are increasingly more willing to be involved in the public sector procedures, the political agents need to design new means to enable the engagement of external agents in the innovation processes of the public sector through innovation techniques (Schmidthuber & Hilgers, 2018).

It is also worth highlighting that open innovation in the private sector is related to the development of physical products or competitive advantages, while the public sector tends to concentrate on non-monetary outcomes (Bommert, 2010; Kivleniece & Quélin, 2011). In general terms, divergences are related to objectives, since the public sector applies its efforts to increase the service performance and the public value (Konsti-Laakso, Hennala, & Uotila, 2008), and the private sector is more focused on maximizing economic and financial outcomes, thus expanding its competitive advantage (Rangan, Samii, & Van Wassenhove, 2006).

## 3  Research method

For this study, a systematic literature review was adopted. It is known as a research technique that follows a well-defined methodology, with properly documented steps, thus ensuring the quality and reliability of the obtained results (Jamshidi, Ghafari, Ahmad, & Pahl, 2012; Reis & Prates, 2011). In comparison with other types of literature reviews (for instance, narrative and scope reviews), the systematic review targets research questions with specific focuses, and with narrow pre-established parameters that are guided by inclusion and exclusion criteria (for instance, topics, settings, types of study). Therefore, it is possible to extract data from only the studies that are included and base a conclusion strictly on the evidence related to the initial research questions (Armstrong, Hall, Doyle, & Waters, 2011; Holeman, Cookson, & Pagliari, 2016).

By following this formal method, with inclusion and exclusion criteria properly established, it is possible to provide a research review replicable with as little bias as possible during the review process of the research findings.

This study was based on the guidelines proposed by Kitchenham (2004) and the procedures carried out were guided by the following perspectives:

1. Characterization of the research terms;
2. Choosing the sources (search engines) in which searches will be conducted;

3. Applying the terms in the search engines; and
4. Selecting primary studies through the inclusion and exclusion application criteria in the research results.

## 3.1 Research questions

Establishing research questions is one of the main differences that distinguish a systematic review from a traditional one. Establishing predefined questions helps to structure the review and guides the reviewing process. This includes the techniques used to identify studies, the critical review of the studies and the analysis of the results.

The objective of this study is to analyze models, methods and approaches on how the public sector has been using the open innovation paradigm to promote innovation, investigating the processes and procedures that facilitate practices and identifying current gaps. For that reason, the following general research question was designed:

1. How does the open innovation process takes place in the public sector?

This general research question was divided into more specific questions, as follows:
RQ1 What open innovation models for governments are there in the literature?
RQ2 What are the approaches used for open innovation in the public sector?
RQ3 Which stakeholders are involved in these initiatives in the public sector?
RQ4 What aspects influence open innovation in the public sector?
RQ5 What is the role of ICTs in open innovation processes in the public sector?

## 3.2 Data sources and search strategy

To plan and execute this review, an exhaustive and wide search of primary studies was carried out. The necessary data to answer the research question(s) were extracted and classified. Figure 1 illustrates the review steps.

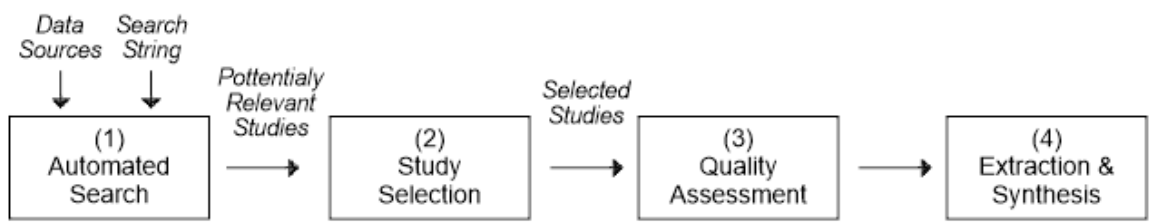

Figure 1. Review steps.

To assure the effectiveness of the search string, two strategies were adopted: (I) a preliminary search was carried out in Google Scholar about research that would involve government and, subsequently, open innovation. This set of articles was used as a basis for the systematic review, in order to compile the main terms that would better represent the research objectives proposed. Two sets of related terms were defined. The main set of terms was related to the approaches in which studies would refer to public agencies in the literature. The second set of terms was related to the paradigm of open innovation. To construct the search string, all of the terms were combined using the boolean operators "OR" and "AND". The reason why these methods were used to create the investigation was to unite the larger possible number of studies that had used the open innovation paradigm in the public sector. Table 1 presents the search terms used in this systematic review. (II) The next step in defining the search strategy was to underline the metadata fields that were more adequate for the search string application. Aiming to reduce the number of publications not relevant to the purpose of this study, it was decided that the search terms would encompass title, abstract and keywords.

Table 1: Search string construction.

| Keyword | Generic search string |
|---|---|
| Government | administrative public sector, central government, city government, country, democratic public, federal government, gov, government, government administration, government agencies, government agency, government innovation, government institution, government institutions, government organization, government organizations, government power, government public sector, government service, governmental public, governments, institution public sector, local government, local governments, municipal government, public administration, public government, public governments, public institution, public institutions, public management, public municipal administration, public organization, public organizations, public power, public sector, public sectors, public service, public urban administration, state, states government |
| Open Innovation | innovation model, open innovation model, public open innovation, open innovation |

With the goal to cover the largest possible quantity of pertinent publications to this theme, a survey on the main electronic libraries that are more widely used in the field of computer sciences was conducted, as listed in Table 2.

Table 2: Digital Libraries.

| Digital Library |
|---|
| ACM Digital Library |
| IEEE Xplore Digital Library |
| Science Direct |
| Springer Link |
| Scopus |
| Tandfonline |

### 3.3 Study selection

First, papers retrieved in the automated search were filtered based on tittle, abstract and keywords. Some results obtained through the strategy mentioned could still be irrelevant to the focus of the research question, even with the terms appearing on the specifics filters. Therefore, a selection of studies had to be carried out, retaining only results relevant in answering the research question. Thus, some inclusion and exclusion criteria were designed to classify the articles during the filtering process. Each potentially relevant article was analyzed by two researchers and reviewed by a third researcher. The conflicting opinions were solved through online meetings. In the end, articles containing results from the same repeated studies and articles were removed to assure that there were no duplications. Google Sheets® was used to register all the steps of the selection process in a group of worksheets, facilitating collaborative work. Publications that met any of the inclusion criteria are selected as primary studies. The inclusion criteria are presented below:

- techniques or methods for using the open innovation paradigm in public organizations;
- experiences or reports on using open innovation in the public sector;
- limiting or encouraging aspects for using open innovation in the public sector;
- practical experience or proposal to use theoretical models for using open innovation in the public sector;

Publications that met any of the following exclusion criteria, as follows, were removed from the review:

- studies that are not related to the investigated topic;
- studies not written in English;
- duplicated studies;
- studies that are not available for download;
- incomplete documents, drafts, interviews, presentation slides, extended abstracts, book chapters or congress proceedings;
- studies that are focused on other types of innovation;
- studies focused on the application of open innovation in a general context (not limited to the context of the public sector);
- studies that were published over 11 years ago (acceptable: 2009 to 2020);
- secondary, tertiary studies, or meta-analysis.

The selection process of primary studies for this review was carried out in July 2021. Consequently, this review included only studies published and indexed before this date. Taking into consideration that the articles of 2021 were in process of development and publication at the moment this study was been developed, they would need more time to be indexed by the search base. Due to that, this review did not contemplate the articles of the year mentioned above, thus taking the replication aspect of these procedures into consideration for future studies.

Only publications written in English were accepted. The year 2009 was selected as the first year, since the topic of open government started to be instrumentalized by the White House on December 8, 2009, thus becoming a mark in the US and, consequently, leading to the creation of the Open Government Partnership in 2011. The availability and standardization of open data was an important step to disseminate open innovation in the public sector. For this reason, we opted to use this year as the starting point, to map out the larger quantity possible of studies on this topic. To properly provide open data is an essential requirement for successful open innovation activities (Thoreson & Miller, 2013).

### 3.4 Overview of included studies

The application of search string engines resulted in an initial set of 4,343 articles. The results were properly registered, while the duplicated studies were removed from the analysis. 2,985 articles were left after this procedure. After these filters, the titles and abstracts from the remaining studies were manually reviewed, removing those entries with titles that indicated they were not relevant for this review. From this manual filter, 192 potential articles were left for the next step. Finally, a full reading of the studies was carried out, applying the inclusion and exclusion criteria defined above. This procedure resulted in 37 studies, available in Appendix A, which represented our final set of primary studies.

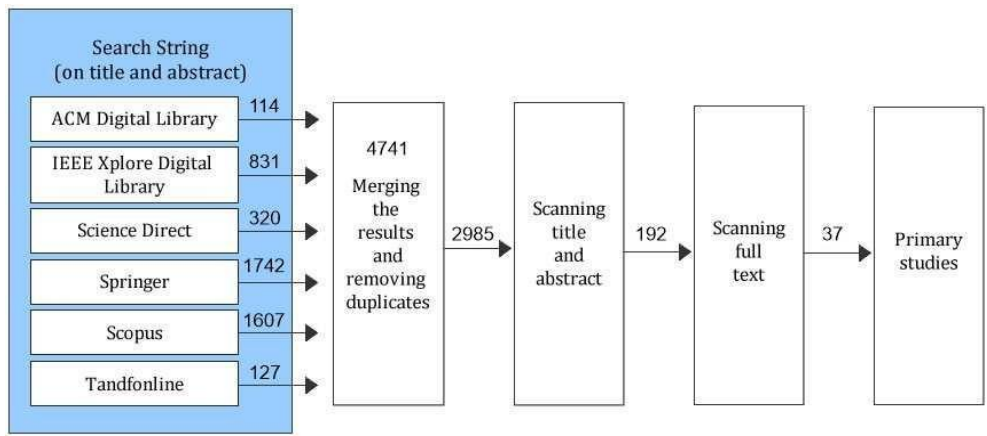

Figure 2. Procedure for identifying primary studies.

The objective of this study is to carry out a systematic analysis of the existent literature in the field of open innovation in the public sector. We discuss here some literature statistics resulting from the conducted systematic analysis. As illustrated in Figure 2, in the year 2009, only one study was shown to be relevant to the purpose of the study. It was found that, although no relevant literature was found in the years 2010 and 2011, results increased significantly in the following years. Even though a series of open data initiatives had already been established, as indicated in the figure, the increased interest in reporting studies on the use of open innovation in government could be potentially connected to the creation of the Open Government Partnership by the end of 2011. The year 2017 presented a larger number of articles related to the purposes of the study. On the other hand, it was found that in 2019 there was a decrease in studies addressing this topic, which was brought back in the year 2020.

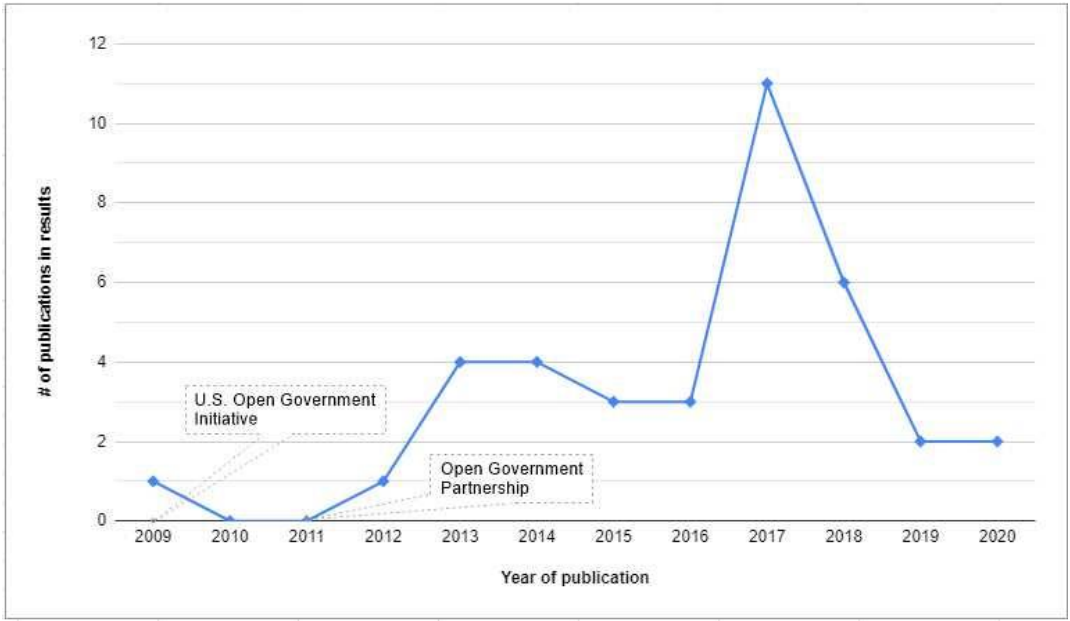

Figure 3: Resulting primary studies by year.

## 3.5 Quality Assessment

After selecting relevant papers, we performed the quality assessment. This procedure was also carried out by two researchers and totally checked by a third one, who also participated in the cases of conflicting evaluations. Table 3 explain the quality assessment criteria.

Table 3: Quality criteria.

| Assessment criteria | Score 0 – 1 | Response |
|---|---|---|
| Is there a clear definition of the study objectives? | | Yes = 1 / No = 0 |
| Is there a clear definition of the reasons for the study? | | Yes = 1 / No = 0 |
| Is there a theoretical basis for the topics of the study? | | Yes = 1 / No = 0 |
| Is there a clear definition of the research question (RQ) and/or the study hypothesis? | | Yes = 1 / No = 0 |
| Is there an adequate description of the context in which the research was carried out? | | Yes = 1 / No = 0 |
| Are appropriate data collection methods used and described? | | Yes = 1 / No = 0 |
| Does the study provide clear answers or justifications for research questions/hypotheses? | | Yes = 1 / No = 0 |
| Is the study providing clearly stated claims with credible results? | | Yes = 1 / No = 0 |
| Does the study provide a clear definition of Government/Open Innovation or a definition of any concept closely related to Government/Open Innovation? | | Yes = 1 / No = 0 |
| Is the study provided justified conclusions? | | Yes = 1 / No = 0 |
| Is there a discussion in the study about limitations of the research provided | | Yes = 1 / No = 0 |

| | Total Quality Score (%) | |
|---|---|---|

## 3.6 Data Extraction and Synthesis

Google Sheets® was used to manage all data extraction, analysis, and synthesis procedure. The following data were extracted: title, abstract, keywords, authors, affiliation, year, journal, publisher, hyperlink of publication and text passages whenever the paper provided answers to the research questions.

To analyze the data, we transcribed passages that answered each research question. We then followed an open coding procedure on these passages, condensing similar codes into thematic categories as in an axial coding procedure. Finally, we tallied the frequency of citations for each category. It is important to note that these frequencies do not reflect the importance of the category, but only how many papers cite them.

## 4 Results

This section presents a discussion of the research results obtained during the literature systematic review.

### 4.1 Overview of the Studies

The conduction of the search string with the sources selected for the development of this research occurred around July 2021 and resulted in a total of 4,741 articles. From that point, a filter was designed aiming to identify articles outside the time range proposed for this study (2009 to 2020). A total of 476 articles were from years before the proposed time of reference, from 1966 to 2008, thus being removed. Regarding the year 2021, the search process in digital libraries happened during that year, hence a total of only 8 studies were found by the digital libraries from that year. Those articles were then removed from the analysis since this number could not represent the totality of studies carried out that year. For this reason, 4,257 articles were left, distributed throughout the years 2009 to 2020, as presented in Figure 3.

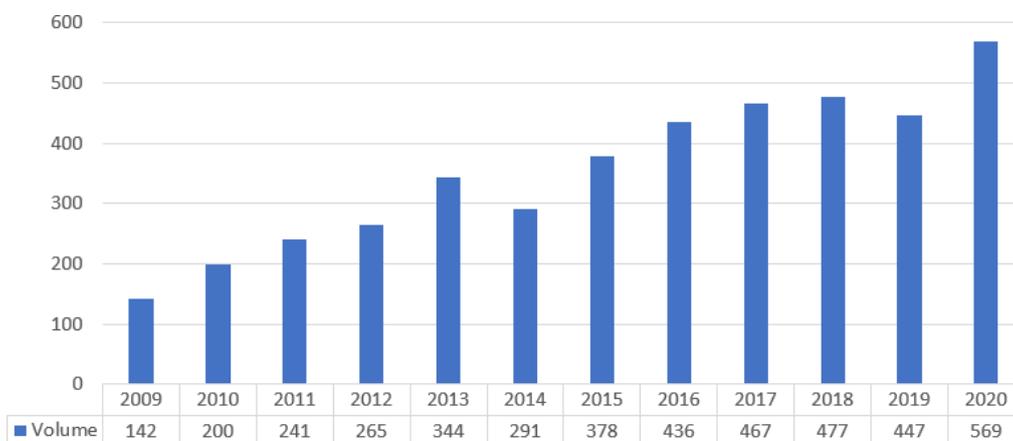

Figure 4: Total of studies found by the search string in the digital libraries.

After a preliminary analysis, the next step was to remove duplicated studies, leading to a total of 2,985 studies. The following step began with the verification of titles and abstracts, excluding those studies with goals not relevant to this review, resulting in a collection of 192 potential studies. In the end, a filter was applied with the exclusion and inclusion criteria of the primary studies. This analysis consisted of the following reading sequence: title of studies, abstracts, conclusions, and the full text, reducing the initial research corpus to 37 studies, defined as our final collection of primary studies.

We discuss here some literature statistics, resulting from the conducted systematic analysis. According to Figure 4, in the year 2009, only one publication was considered relevant to the purpose of this study. It was found that, although no relevant literature was found in the years 2010 and 2011, results increased significantly in the following years. Even though a series of initiatives of open data had already been established, as indicated in the image, the increase in the interest in reporting studies on the use of open innovation in government could be potentially connected to the creation of the Open Government Partnership by the end of 2011. The year 2017 presented the larger number of articles related to the purposes of the study.

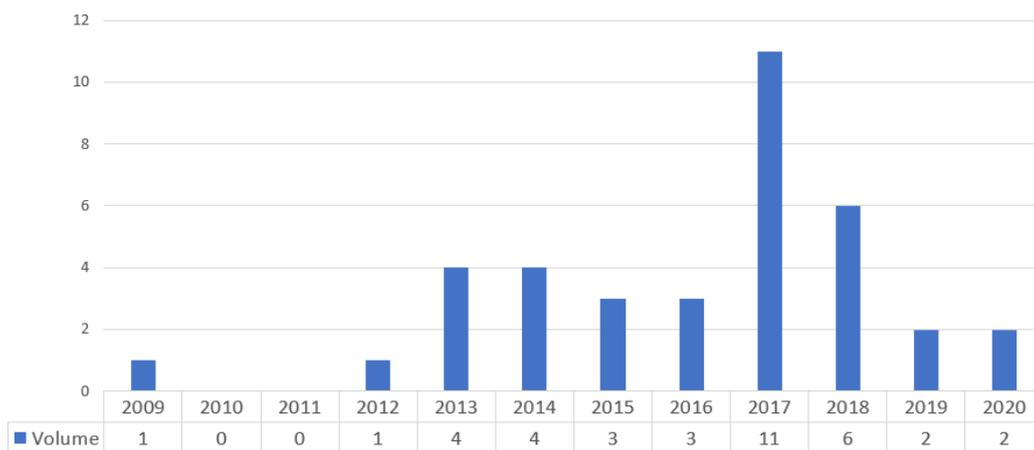

Figure 5: Total of studies selected by the inclusion and exclusion criteria.

Through the inclusion and exclusion criteria, studies that were related to results referent to the use of open innovation in the private sector were removed. Primary studies that would only make references or quotes about the theme, but didn't address the process, technique, approach or report of open innovation in the public sector were also excluded. The English language was chosen as a criterion for primary studies since the most relevant journals and events of the field publish research mainly in English.

Springer represented the larger number of articles in the preliminary search (38.05%), but only one article met the selection criteria for the present study. On the other hand, Scopus had the second larger number of articles (32.82%) and presented the largest percentage of acceptance (56.76%). Table 4 synthesizes these data.

Table 4: Presentation of identified and selected primary studies

| Total primary studies | | | Total selected studies | | |
|---|---|---|---|---|---|
| Digital library | Total | % | Digital library | Total | % |
| ACM Guide | 106 | 2.49% | ACM Guide | 3 | 8.11% |
| IEEE | 754 | 17.71% | IEEE | 3 | 8.11% |
| Science Direct | 267 | 6.27% | Science Direct | 8 | 21.62% |
| Scopus | 1397 | 32.82% | Scopus | 21 | 56.76% |
| Springer | 1620 | 38.05% | Springer | 1 | 2.70% |
| Tandfonline | 113 | 2.65% | Tandfonline | 1 | 2.70% |
| **Total** | **4257** | **100**% | **Total** | **37** | **100**% |

From the selected primary studies, it was possible to identify 94 different authors that contributed to the theme at this period. Ines Mergel and Mila Gasco-Hernandez were the main contributors with four articles each. Besides them, other authors who also stood out were Aggeliki Androutsopoulou, Unai Aguilera, Mikel Emaldi, Euripidis Loukis, Diego López-de-Ipiña, Paulo Henrique De Souza Bermejo, Jorge Pérez-Velasco, Sandoval-Almazan R. and Yannis Charalabidis. These data are summarized on Table 5.

Table 5: Authors who contributed to the theme

| Author | References | # |
|---|---|---|
| Ines Mergel | I2162, I2528, I2765, I3797 | 4 |
| Mila Gasco-Hernandez | I3395, I3424, I3835, I4151 | 4 |
| Aggeliki Androutsopoulou | I3461, I3552 | 2 |
| Unai Aguilera | I3191, I3553 | 2 |
| Mikel Emaldi | I3191, I3553 | 2 |
| Euripidis Loukis | I3461, I3461 | 2 |
| Diego López-De-Ipiña | I3191, I3553 | 2 |
| Paulo Henrique De Souza Bermejo | I2487, I2749 | 2 |
| Jorge Pérez-Velasco | I3191, I3553 | 2 |
| Sandoval-Almazan R. | I3424, I3442 | 2 |
| Yannis Charalabidis | I3461, I3552 | 2 |

## 4.2 Study Focus

From the selected primary studies, it is possible to classify, based on their objectives, the focus of the study into four distinct objectives:

1. **Experience report:** case studies on initiatives of open innovation in the public sector (29.73%);

2. **Approach proposal:** frameworks, platforms, processes and models with the objective to assist the open innovation process in the public sector (27.03%);

3. **Understanding the phenomenon:** the use of open innovation through crowdsourcing technique, the role of intermediaries and exploring factors that promoted or hindered the implementation of open innovation (24.32%);

4. **Theoretical Reflection** the importance of involving intermediaries, future directions on open innovation in government, overcoming cultural limitations to increase the level of collaboration between public agencies and companies, and ICTs as support for the process of open innovation (18.92%).

Table 6 presents primary studies that are linked to these classifications and their respective percentages.

Table 6: Classification of primary studies

| Classification | References | # | % |
| --- | --- | --- | --- |
| Experience reports | I1473, I2162, I2456, I2487, I2749, I3442, I4151, I2093, I2477, I4575, I4154 | 11 | 29.73% |
| Approach proposal | I2073, I2606, I2877, I3047, I3191, I3358, I3461, I3553, I3791, I4302 | 10 | 27.03% |
| Understanding the phenomenon | I1887, I3395, I3571, I3627, I3797, I2528, I3502, I3908, I3606 | 9 | 24.32% |
| Theoretical reflection | I3424, I3437, I3438, I3552, I3835, I2117, I2765 | 7 | 18.92% |

Most primary studies are directly interested in reporting experiences (29.73%) of open innovation initiatives in the public sector through case studies of one or several initiatives. Another recurring topic is the proposal (27.03%) of approaches or processes aiming to assist the public sector in adopting open practices. Among these approaches, the studies propose frameworks (40%), platforms (33%), conceptual models (22%) and processes (11%). Some studies are set to understand the phenomenon (24.32%) from the use of intermediaries (living labs, urban labs, crowdsourcing platforms) among other factors that make it difficult to implement open innovation in the public sector and evaluate the main opening practices by public agencies. Finally, other studies propose to reflect (18.92%) the role of intermediaries, regarding on what are the future directions of open innovations in the public sector, the appropriate combination for the use of ICTs in government, and to understand how cultural limitations can increase the level of collaboration.

### 4.3 Quality of Primary Studies

As described previously, each study was assessed independently by two researchers according to six possible quality criteria, and eventual conflicts were solved by a third researcher. Table 7 summarizes the quality assessment, in which the primary studies are grouped according of the quality score. Only two studies (I2477 and I4302) received the maximum score. On the other hand, five studies (I2456, I2606, I3553, I3191, I3835) scored below 50%.

Table 7: Quality scores.

| Score | Papers |
| --- | --- |
| 100% | I2477, I4302 |

| 90,91% | I2765, I3358, I3395, I3438, I3797, I4151 |
| 81,82% | I1473, I3442, I4575 |
| 72,73% | I2073, I2093, I2487, I2528, I3424, I3461, I3502, I3908, I4154 |
| 63,64% | I2117, I2162, I3047, I3437, I3571, I3627 |
| 54,55% | I1887, I2749, I2877, I3552, I3606, I3791 |
| 45,45% | I2456, I2606, I3553 |
| 36,36% | I3191, I3835 |

## 4.4 Answers to the Research Questions

To summarize the data obtained from the systematic review, the results are presented based on the research questions as highlighted in the next subsections 3.1.

### 4.3.1 [RQ1] What open innovation models for governments are there in the literature?

The systematic review identified two studies that proposed a conceptual model to support the innovation process in the public sector. Besides these models, it was possible to map out a group of processes and strategies of open innovation for public agencies. Table 8 lists these findings.

Table 8: Summarizing models, processes and strategies identified.

| Proposition | References |
|---|---|
| Innovation Model | I3791, I4302 |
| Open Innovation Process | I2765 |
| Open Innovation Strategies | I4154 |

The innovation model proposed by the study [I3791] involves the 4 types of interested parties (government, companies, citizens and researchers) performing an active role throughout the whole lifecycle, since the identification of needs to a successful solution. The study reports a wide range of citizens engaged in this co-creation/co-production: common citizens (students, neighborhood residents), specialists from several fields (artists, creative people, academics, entrepreneurs, technicians, activists, business representatives – manufacturers, tradesmen, retailers, consultants), research communities and other agencies from the public sector.

The study [I4302] proposes the development of an innovation model that seeks to assist government agencies to identify, design and conceptualize ideas in order to promote improvements and explore new services in a more efficient and effective manner. The proposed model focuses in three main areas: (1) the type of innovation model, (2) the steps of the process as a whole and (3) the tools associated with each step. The model can be applied internally in organizations. It consists in a structure of six distinct steps, including a filter, two decision gates and iterative loops that are employed in some stages to obtain decisions before advancing to the next stage. These iterations allow the model to take into consideration information from several internal and external sources.

Based on theoretical reflections and a case study about a crowdsourcing platform, the study [I2765] identified that after the government agencies defined their public management problem, they go through four distinct phases in an open innovation process: (1) generation of ideas, (2) incubation, (3) validation and (4) unveiling the selected solution and implementing (internally) the winning idea. This group of processes was adopted by the studies [I3442] and [I3552]. The former incorporated the proposed phases to support a process of open innovation that was idealized through a case study. The latter reflected on how the proper combination of ICT tools can support the implementation of open innovation practices in the public sector.

Based on a number of initiatives, the study [I4154] identified three open innovation strategies. The first strategy consists of a basic collaboration step, where the city involves other agents (city residents or a specialist) with the objective to cocreate solutions. In the second proposed strategy, the city takes on the role of the financing agency and infrastructure facilitator for the process, aiming to expand open innovation for the public sector. The third open innovation strategy is observed in the initiatives in which all interested parties are intimately involved throughout the collaboration process. The three open innovation strategies proposed by the study are inspired in the inbound, outbound and coupled flows.

### 4.3.2 [RQ2] What are the approaches used for open innovation in the public sector?

From the initial group of primary studies, twenty-four articles (64.86%) provided answers on the main approaches used for open innovation in the public sector. The primary studies revealed a total of ten approaches, which are summarized in Table 9 and highlighted below:

- Web Platforms (24.32%): are commonly proposed to subsidize the innovation process through platforms, forums, proposing ideas, sending suggestions, complaints, and establishing communication and collaboration among those involved in the process;

- Social Networks (13.51%): Monitoring social networks and integration with network communities were the most reported types of approach. These initiatives had as their main goal to involve external agents in the innovation process through these interactions. Examining the reports, it was possible to verify that the government agencies that wanted to have open innovation initiated their experimentations using social networks;

- Crowdsourcing (10.81%): the use of the crowdsourcing technique was also reported as an open innovation approach for the public sector. It consists of a challenge, posted online, and an award offered for the best answer to the challenge. Among the platforms related to the studies, it is possible to verify Challenge.gov as a well-successful experience in the US, with the main goal to support federal agencies, while promoting and expanding the use of award competitions to identify innovative solutions for critical questions;

- Open data (10.81%): Opening data and making it available in platforms allows for public consultations. Several government agencies have been promoting open innovation through platforms that have data dictionaries and APIs to enable interaction;

- Gamified Platforms (8.11%): some studies report the use of gamification as a proposal for open innovation, especially in the preliminary stages of interaction between citizens and their city;

- Hackathons (8.11%): the use of programming marathons, known as hackathons, became a viable alternative to open innovation for agencies of the public sector. The experiences reported on hackathons are directly associated with the opening of data, used as a necessary input for creating and proposing technological solutions;

- Crowdstorming (5.41%): distributed and collaborative methods of brainstorming were also reported as an open innovation approach in the public sector. The crowdstorming approach consists of uniting the largest number of ideas about a specific domain or topic. This approach may be found in initiatives such as the Office of Social Innovation and Civic Participation in the US (Office of Social Innovation and Civic Participation, 2022);

- Technology (5.41%): some articles present reports on how technological features and tools can support and promote the implementation of open innovation practices in the public sector.

The results mentioned are summarized in Table 9.

Table 9: Approaches of open innovation used in the public sector.

| Approach | References | # | % |
|---|---|---|---|
| Web Platform | I1473, I2093, I2606, I2877, I3191, I3358, I3553, I3797, I3908 | 9 | 24.32% |
| Social Media | I2073, I2749, I3437, I3461, I3502 | 5 | 13.51% |
| Crowdsourcing | I2477, I2765, I3908, I4575 | 4 | 10.81% |
| Open data | I2477, I3358, I3606, I3908 | 4 | 10.81% |
| Gamified Platforms | I2877, I3047, I3442 | 3 | 8.11% |
| Hackathons | I2477, I4151, I4575 | 3 | 8.11% |
| Crowdstorming | I2487, I3797 | 2 | 5.41% |
| Technology | I3552, I3835 | 2 | 5.41% |

### 4.3.3 [RQ3] Which stakeholders are involved in these initiatives in the public sector?

Based on the primary studies, thirty articles (81.08%) provided information on the main stakeholders and their respective roles in the open innovation processes in the public sector. In total, twenty-four stakeholders were catalogued, and among the selected studies, citizens, companies, intermediaries and government were highlighted. Table 10 presents the data regarding stakeholders.

Table 10: Stakeholders involved with open innovation initiatives in the public sector.

| Stakeholder | References | # |
| --- | --- | --- |
| Citizens | I1887, I2117, I2162, I2477, I2528, I2606, I2749, I2877, I3047, I3191, I3358, I3395, I3442, I3461, I3502, I3553, I3791, I3908, I4151, I4154, I4575, I2093, I2456, I3571 | 24 |
| Private Companies | I1887, I2117, I2456, I2477, I3047, I3191, I3395, I3553, I3791, I4154, I4575, I2093, I1473, I2749, I2162 | 15 |
| Intermediaries | I1887, I2117, I2162, I2456, I2477, I2528, I3358, I3395, I3424, I3438, I3553, I3627, I3797 | 13 |
| Governments | I1473, I2093, I2117, I2477, I2606, I3553, I3791, I4575, I3571, I4154, I4151 | 11 |
| Universities | I3047, I3395, I3553, I3908, I4154, I4575 | 6 |
| Internal Staff | I1473, I2487, I2749, I2877, I3191, I3791 | 6 |
| Researchers | I2456, I2487, I3191, I3791, I4575 | 5 |
| Civil Society | I2477, I3358, I3627 | 3 |
| NGOs | I2477, I4154 | 2 |
| Non-profit Entities | I1887, I2162 | 2 |
| Other public institutions | I2456, I2477 | 2 |
| Startups | I2117, I2456 | 2 |
| Project Committees | I1473, I2877 | 2 |
| Citizen Staff | I1887, I4151, I2162 | 2 |
| Innovators | I2456 | 1 |
| Joint-venture | I2477 | 1 |
| R&D Labs | I2456 | 1 |
| Moderators | I2606 | 1 |
| Thinkers | I2456 | 1 |
| Citizen representatives | I2093 | 1 |
| User Committees | I2877 | 1 |
| Sector Syndicates | I1887 | 1 |
| Foundations | I4154 | 1 |

The citizens have essential roles in the open innovation processes and are used as the primary input for supplying ideas and solutions to the public field. Among these roles, the studies classify the citizen as a predominant source of external innovation. In one of the studies, the city employed a citizen as an internal facilitator, or 'agent citizen', aiming to encourage other citizens to engage in the public institution initiatives, while supplying city managers and other citizens with information. There were also reports of open innovation processes in which the public agency defined which would be the skills citizens should possess to meet their needs, especially developers, scientists and students. Table 11 summarizes the main roles attributed to citizens.

Table 11: Citizens set of roles in the processes of open innovation.

| Stakeholders | Roles | References | # |
|---|---|---|---|
| Citizens | External source of innovation | I1887, I2093, I2117, I2162, I2477, I2528, I2606, I2749, I2877, I3047, I3191, I3358, I3395, I3442, I3461, I3502, I3553, I3791, I3908, I4151, I4154, I4575 | 21 |
| | Facilitator | I2093 | 1 |
| Citizen Developers | External source of innovation | I2117, I2477, I3571, I4154 | 4 |
| Citizen Scientists | External source of innovation | I2456 | 1 |
| Citizen Students | External source of innovation | I3908 | 1 |

The literature also highlights the involvement of private companies in collaboration with public agencies during the processes of open innovation. The primary studies brought to light a total of 8 roles attributed to companies. Among those roles, private companies are involved with the objective to provide technical support and collaborating with the innovation process, thus being external sources of innovation. Besides that, companies also are involved in the process of providing solutions for challenges and taking the facilitator role during the process. Table 12 presents their main roles.

Table 12: Companies set of roles in the processes of open innovation.

| Stakeholders | Roles | References | # |
|---|---|---|---|
| Private Companies | Opening data | I2477 | 1 |
| | External source of collaboration | I1887, I2117, I2456, I2477, I3047, I3191, I3395, I3553, I3791, I4154, I4575 | 11 |
| | Facilitator | I4154 | 1 |

| | Funder | I1473 | 1 |
| | Supplier | I2749 | 1 |
| | Solution provider | I1887, I2162 | 2 |
| | Establishing policies on data supply | I2477 | 1 |

Intermediaries are considered important supporting roles in the open innovation process in the public sector. In total, thirteen articles reported eight types of open innovation. From the number of roles attributed, it is possible to highlight the use of crowdsourcing platforms, Living Labs and Urban Labs. Crowdsourcing platforms are commonly used as mechanisms for obtaining ideas, knowledge and solutions for the challenges institutions face. These online platforms create awareness of unsolved challenges and unite citizens in a competitive scenario set to solve problems online. One of the most used crowdsourcing platforms for conducting case studies is Challenge.gov. Living Labs are also set as intermediaries of open innovation and usually collaborate through lectures, and by giving support and feedback to the public sector. Its activities are related to building solutions for the public scenario, creating a space in favor of innovation, and expanding the public knowledge on the use of APIs and open data. At last, Urban Living Labs contribute directly to the process of open innovation by providing physical structures to citizens, recruiting and maintaining a community of developers willing to participate in innovation strategies (e.g. hackathons, and development contests) and acting as agents of change in the city halls' organizational structures. Table 13 lists the types of intermediaries related to these studies.

Table 13: Types of intermediaries of open innovation in the public sector.

| **Types** | **References** | **#** |
|---|---|---|
| Crowdsourcing Platforms | I1887, I2162, I2528, I3358, I3797 | 5 |
| Living Lab | I2117, I3395, I3424 | 3 |
| Innovation Labs | I2456 | 1 |
| Urban Living Labs | I2477, I3627 | 2 |
| Fab Labs | I3395 | 1 |
| Innovation Labs | I3424 | 1 |
| Centres of Public Research | I3438 | 1 |
| Web Platform | I3553 | 1 |

Lastly, the eleven primary studies revealed a total of ten roles taken by government. Besides relying on the assistance of external agents, governments also make use of their own resources and collaborators to promote innovation along with citizens. Besides that, they also usually take the role of facilitators and funders of open innovation. Studies also show that government needs to be involved and to provide all the necessary support for citizens to conduct the process of innovation. Table 14 lists these roles. The primary studies revealed other nineteen stakeholders, besides the four already mentioned, and their respective roles, presented in Table 15.

Table 14: Government set of roles in the processes of open innovation.

| Stakeholders | Roles | References | # |
|---|---|---|---|
| Government | Internal source of innovation | I1473, I2093, I2117, I2477, I2606, I3553, I3791, I4575 | 8 |
| | Innovation Facilitator | I4151, I4154 | 2 |
| | Funder | I1473, I4154 | 2 |
| | Support | I2117, I4154 | 2 |
| | Supplier of Open Data and APIs | I2117, I2477 | 2 |
| | Assist | I3571 | 1 |
| | Purchasing services | I2117 | 1 |
| | Knowledge / Specialist | I4154 | 1 |
| | Collaboration Network | I2117 | 1 |

Table 15: Other Stakeholders and their roles in the use of open innovation in the public sector.

| Stakeholders | Roles | References | # |
|---|---|---|---|
| Universities | External source of innovation | I3047, I3395, I3553, I3908, I4154, I4575 | 6 |
| Internal Staff | Internal source of innovation | I1473, I2487, I2749, I2877, I3191, I3791 | 6 |
| Researchers | External source of innovation | I2456, I2487, I3191, I3791, I4575 | 5 |
| Civil Society | External source of innovation | I2477, I3358, I3627 | 3 |
| NGOs | External source of innovation | I2477, I4154 | 2 |
| Non-profit Entities | Solution provider | I1887, I2162 | 2 |
| Other public institutions | External source of innovation | I2456, I2477 | 2 |
| Startups | External source of innovation | I2117, I2456 | 2 |
| Project Committees | Assisting | I1473, I2877 | 2 |
| Citizen Staff | Solution provider | I1887, I2162 | 2 |
| | Facilitator | I4151 | 1 |
| Innovators | External source of innovation | I2456 | 1 |
| Join-ventures | Encouraging | I2477 | 1 |
| R&D Labs | External source of innovation | I2456 | 1 |
| Moderators | Facilitator | I2606 | 1 |
| Thinkers | External source of innovation | I2456 | 1 |
| Citizen reps | Facilitator | I2093 | 1 |
| User Committees | Planning Challenges | I2877 | 1 |
| Sector Syndicates | Solution provider | I1887 | 1 |
| Foundations | External source of innovation | I4154 | 1 |

### 4.3.4 [RQ4] What aspects influence open innovation in the public sector?

Based on the primary studies, it was possible to verify the existence of a group of thirty aspects that are involved during the open innovation process. From this, nineteen aspects (63.33%) were identified as possible aspects that promote open innovation. Table 16 consolidates these promoting aspects.

Table 16: Promoters in the open innovation processes in the public sector.

| Factor | References | # |
|---|---|---|
| Collaboration | I2073, I2162, I2456, I2477, I4575 | 5 |
| Opening Data | I1887, I2117, I4575 | 3 |
| Technologies | I2456, I2477 | 1 |
| Laid-back Environment | I3047 | 1 |
| External Actors | I4151 | 1 |
| Citizens Awareness | I2117 | 1 |
| Direct Contact | I2093 | 1 |
| Disclosing Challenges | I2528 | 1 |
| People Empowerment | I4151 | 1 |
| Contracts Facility | I1473 | 1 |
| External Facilitator | I2093 | 1 |
| Financing | I1473 | 1 |
| Gamification | I3047 | 1 |
| Intermediaries | I2117 | 1 |
| Pertinent Legislation | I2528 | 1 |
| New Challenges | I2117 | 1 |
| Pecuniary Rewards | I2477 | 1 |
| Solving Problems | I2528 | 1 |
| Transparency | I2073 | 1 |

Based on the combined promoting factors, it was noted that collaboration and opening data presented a higher consistency. Collaboration is viewed as a fundamental aspect of innovation and commonly surges from integration between government, civil society and companies, with the objective to integrate external partners and discuss solutions when facing problems in the public sphere. Another promoting factor is the availability of a collection of open-format data in government platforms. Initiatives of open data promote a whole innovation ecosystem, bring new business for private companies and enable inputs for the intermediaries of open innovation (Living Labs, Urban Labs, among others).

On the other hand, eleven aspects (36.67%) were classified as barriers to the open innovation process in the public sector. Table 17 lists the barriers identified.

From this collection of barriers to open innovation in the public sector, the ones with the higher concentration are: bureaucratic factors, legal barriers, limitations in the process and resistance to opening data. The bureaucratic aspects are closely related to the condition of the public sector's legislation for acquiring products and services. The results show that the traditional process of innovation is highly regulatory and follows rigid rules and regulations, while open innovation needs more freedom to operate within public agencies. Another key aspect is the legal barrier, as every and any procedure to be carried out in the public sector needs a whole set of regulations and approvals by other departments, which is one of the main barriers reported. Other limitations are the procedures adopted to innovate. According to the studies, the agencies that use innovation strategies such as hackathons are faced with insufficient final results for the proposed problem. Besides that, the next steps after the competition weren't clear or objective, leading to failures by the public agencies that intended to implement those actions. Lastly, another reported aspect is the resistance some agencies impose to opening their databases. There were reports in which the process of opening data was considered a 'burden' and, therefore, only a few agencies gave priority to opening data.

Table 17: Barriers in the open innovation processes in the public sector.

| Factor | References | # |
| --- | --- | --- |
| Bureaucracy | I1473, I1887, I2162, I3797 | 4 |
| Legal Barriers | I1887, I2162, I3797 | 3 |
| Process limitation | I4151, I4575 | 2 |
| Resistance to Opening Data | I2162, I2477 | 2 |
| Internal Conflicts | I2477 | 1 |
| Lack of Feedback | I2073 | 1 |
| Lack of Transparency | I2765 | 1 |
| Integrating Intermediaries | I2477 | 1 |
| Political Mandate | I3797 | 1 |
| Privacy of Personal Data | I2073 | 1 |
| Technical Restrictions | I1887 | 1 |

### 4.3.5 [RQ5] What is the role of ICTs in open innovation processes in the public sector?

Seventeen studies provided answers to RQ5. The role of ICTs in the processes of open innovation in the public sector was extracted from these articles. Table 18 lists the roles defined by the studies.

ICTs are present in the intermediation of the innovation process, in the support of collaborative activities, in open data, availability of services, communication and monitoring of social networks. The roles of intermediation refer to the use of web platforms as innovation intermediaries, assisting to achieve solutions and, consequently, identifying and solving problems. The role to support collaborative activities was emphasized in the literature. The studies highlighted the usage of ICTs allowing group discussions, sending comments and evaluating submitted ideas, and the possibility of visualizing the ideas sent by other citizens, voting, debating ideas, and online forums, among other aspects. The availability of open data in interoperable format has an essential role. Technology has also an essential role in presenting services and products. The studies highlighted that web- sites and mobile apps are used for presenting the catalogue of offered services, disclosure of databases and web platforms for displaying ideas. Another essential role in diffusing information is communication. Studies report that technology has an important role in the public transmission of open innovation calls. Lastly, monitoring social networks has been one of the pioneer strategies to engage citizens and government in favor of innovation.

Table 18: Role of Information and Communication Technologies in the processes of open innovation in the public sector.

| Role | References | # |
|---|---|---|
| Intermediate the innovation process | I1887, I2162, I2487, I2528, I2749, I2877, I3191, I3358, I3437 | 9 |
| Support collaborative activities | I1473, I2456, I2487, I2749, I2877, I3191, I3358, I3908 | 8 |
| Open Data | I2117, I2456, I2477, I3358, I3791, I3908 | 6 |
| Services availability | I2456, I2477, I2749, I3437, I3908 | 5 |
| Communication Possibility | I1887, I2162, I2487, I2528, I3437 | 5 |
| Monitoring Social networks | I2073, I2749, I3437, I3461 | 4 |
| Processes Facilitator | I1473, I2456 | 2 |
| Development of new technologies | I2117 | 1 |
| APIs availability | I2117 | 1 |
| Gamification of the innovation process | I2749 | 1 |

## 5 Discussion

The results of this review suggest that open innovation can benefit Government institutions as a strategy to address challenges and promote economic development. Key stakeholders of open innovation in government include citizens, private companies, innovation intermediaries, other governments, universities, employees, non-governmental and social organizations. Collaboration among these actors is essential for co-creating innovative solutions for public problems, which can also increase government transparency and accountability.

Open innovation in the public sector can lead to changes in the way the government operates, such as using web platforms and social networks to engage citizens in decision-making, crowdsourcing initiatives, hackathons, and crowdstorming to solve complex challenges, and opening data for new business opportunities, innovation, and collaboration. However, government institutions need to manage and implement these initiatives properly, ensuring privacy and security of personal data and handling conflicts and challenges that may arise during the innovation process.

Implementing open innovation may require cultural and organizational changes, and public managers must be able to identify opportunities and establish clear objectives and performance indicators. Open innovation may also require greater flexibility and adaptability from public organizations to quickly adjust to changes in citizens' and market needs. Citizens can play diverse roles in open innovation processes, and technology plays a crucial role in facilitating collaboration and co-creation among different actors.

This review can guide the development of public policies that promote open innovation and highlights the importance of cross-sector collaboration and co-creation for solving complex problems. Public organizations should be open to learning from other successful open innovation initiatives.

### 5.1 Practical and Managerial implications

Government institutions should focus on promoting collaboration, opening data, and leveraging ICTs to overcome barriers and enhance open innovation processes in the public sector. However, there are practical implications for open innovation in the public sector, such as:
- Proper articulation with institutions that play the role of intermediaries- it is important to involve intermediaries that are able to handle many aspects to help promote open innovation in the public sector. Governments should rely on partner institutions that can support them in establishing partnerships with civil society and companies in order to integrate such external partners and discuss solutions to public problems;
- Appropriation of Information and communication technologies- ICTs play a key role in open innovation processes in the public sector. Governments should leverage the use of web platforms as well as use technology to support collaborative activities such as group discussions, commenting, and evaluating ideas. Governments should also ensure the availability of open data in an interoperable format, and promote the use of websites and mobile apps for presenting the catalog of services and products. Communication through technology and monitoring social networks can also help in diffusing information and engaging citizens and government in favor of innovation;
- Open data- Open data and ICT play an essential role in supporting innovation. It helps promote innovation, expand public knowledge, and enable collaborative activities. Making data available on platforms allows for public consultations and supports open innovation through the use of APIs and direct manipulation of raw data. Initiatives of open data also promote an innovation ecosystem, which creates new business opportunities articulated by open innovation intermediaries such as Living Labs and Urban Labs;
- Overcoming obstacles- Bureaucratic factors, legal barriers, limitations in the process, and resistance to opening data are major barriers to open innovation in the public sector. Governments should work towards reforming their legislation to make the acquisition of products and services more flexible and less regulatory. There should also be clear and objective next steps after innovation competitions to avoid failures by public agencies. Furthermore, efforts should be made to address resistance to opening data, which can be achieved by promoting the benefits of open data to public agencies.

Open innovation in government institutions can provide a more efficient and effective way for the public sector to identify, design, and implement innovative solutions by involving various stakeholders in the innovation process. The use of open innovation models and approaches is part of strategic moves that can help government agencies to promote improvements and explore new services. There are managerial implications involving the following aspects:
- Open innovation models- The choice of an appropriate model for a Government institution's needs is key to an open innovation strategy. In this review, two conceptual models were identified, one involving four types of interested parties (government, companies, citizens, and researchers) playing an active role in the entire innovation process, while the other proposes a six-step structure to assist government agencies in identifying, designing, and conceptualizing ideas to promote improvements and explore new services;
- Open innovation approaches- The decision on which approaches to put into practice is also strategic. There were several ICT-centered approaches identified, including web platforms, social networks, crowdsourcing, open data, gamified platforms, hackathons, crowdstorming, and technology. These approaches can be used to subsidize the innovation process, involve external agents in the innovation process, offer awards for innovative solutions, open data for public consultation, and unite the largest number of ideas about a specific domain or topic;
- Stakeholders involvement- The approach on how and which stakeholders to involve in an open innovation process concerns another strategic choice that can impact managers and decision-makers. The stakeholders involved in open innovation initiatives in the public sector include government agencies, citizens (common citizens, specialists from several fields, and research communities), companies, researchers, and other agencies from the public sector. For instance, these two strategy options may go in different directions: focus on having citizen engagement as a primary goal or create new businesses and foster the local innovation ecosystem.

## 5.2 Conceptual Model of Open Innovation applied in the Public Sector

After analyzing the data obtained from the research questions, we constructed a conceptual model (Figure 6) that showcases the main components that surfaced during our literature review. The central point of the model is the Government institution, which we assume adopts an innovation model (RQ1) and selects one or multiple approaches (RQ2) that can be applied at different stages of the innovation process to acquire decisions before moving on to the next phase. These approaches commonly revolve around Information and Communication Technologies (ICT), like a Web platform, Crowdsourcing or Hackathons. The ICTs identified in the model have diverse roles (RQ3), including facilitating the innovation process, providing services, or acting as an API.

The innovation model identified in the answer to RQ1 involves the participation of multiple stakeholders. Furthermore, our findings from RQ5 suggest that there are barriers (e.g., bureaucracy) that hinder open innovation, but there are also factors that promote it (e.g., collaboration).

Overall, this conceptual model can serve as an initial framework for understanding the key elements of open innovation in the public sector and help guiding future research in this area. By identifying the factors that promote and hinder open innovation, Government institutions can make informed decisions when adopting innovation models and selecting ICT-based approaches to drive innovation forward.

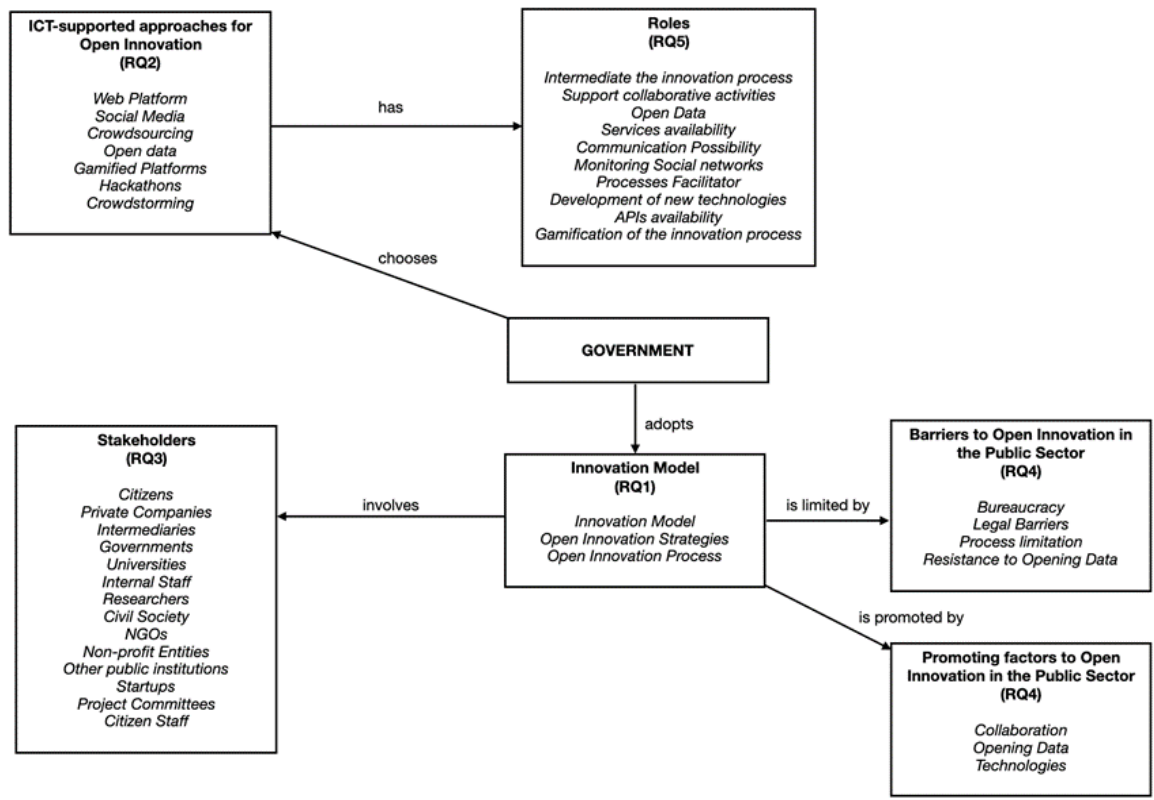

Figure 6: Conceptual model for Open Innovation in the Public sector.

### 5.3 Limitations of this Review

The most common limitations in a systematic review are potential biases introduced during the selection process and data extraction errors. These are the primary potential constraints of this study. All review stages were completed in pairs, disputes in the selection process were resolved by a third party or in consensus sessions, and all justifications for including and excluding studies at each point were documented.

### 6 Summary and conclusions

This review analyzed a total of 4,741 studies, from which 37 provided answers to the research questions. A total of two conceptual models of open innovation were identified, regarding the public sector. Besides those models, other studies were also identified for proposing processes and steps to implement open innovation in the public sector. Among the main approaches of open innovation, web platforms presented a higher rate of reports among the selected studies. Regarding web platforms, crowdsourcing platforms stood out. When considering the main stakeholders involved in the open innovation initiatives, citizens represented the larger intersection among the primary studies, confirming what is portrayed by the literature, with regard to the involvement of citizens in these initiatives. Among the main influencers, collaboration is one of the main drivers of open innovation in contrast to bureaucracy and legal aspects which are the main barriers. As for the main role performed by technology, the feature of supporting the open innovation process is the most common one, as pointed out by many studies.

These results present a general perspective of how the literature has reported the process of open innovation in public agencies and the main elements involved in these processes. The results confirm that the usage of the innovation paradigm in the public sector is still a new field and it is still in an exploratory phase, needing more studies about the systematization by the scientific community. Besides, there are barriers and challenges imposed against the innovation processes, since the public sector has several guidelines and rules that prevent the proper usage of open innovation mechanisms, leading to a need for reflection on which procedures and methods need to be adopted in the public sphere.

The implications of this systematic review point toward the need to structure models and processes of open innovation in government that could promote a better articulation between stakeholders, technologies, platforms and processes. When looking deeper into these initiatives there are a few explicit models– either intentional or unintentional – of open innovation being applied to the public sector. This indicates that most initiatives are commonly carried out without a structured process or a concise methodology, being organized in a rather exploratory manner. In order to be adopted or replicated, such models and processes need to be systematically mapped. It thus reveals a gap to be explored in research.

To expand the understanding of this topic, as well as to promote future works, we propose future studies to include the perspective of the main stakeholders who are engaged in the open innovation initiative in the public sector. We intend to map the models or processes they have been following to promote such initiatives and contrast them with what is found in literature.

**Appendix A    Selected primary studies**